\documentclass[11pt]{article}

\PassOptionsToPackage{numbers,sort&compress}{natbib}
\usepackage{acl}

\usepackage{times}
\usepackage{latexsym}
\usepackage{adjustbox}
\usepackage{tabularx}
 \usepackage{booktabs}
\usepackage[T1]{fontenc}
\usepackage{tikz}
\usetikzlibrary{positioning,fit,arrows.meta,patterns,calc}

\usepackage[utf8]{inputenc}

\usepackage{microtype}

\usepackage{inconsolata}

\usepackage{graphicx}

\pdfoutput=1
\usepackage{pgfplots}
\pgfplotsset{compat=1.18}
\usepackage{xcolor}

\usepackage{tikz}
\usetikzlibrary{shapes.geometric, arrows}

\usepackage{pgfplots}
\pgfplotsset{compat=1.18}
\usepackage{pgfplotstable}

\tikzstyle{startstop} = [ellipse, minimum width=2cm, minimum height=1cm,text centered, draw=black, fill=white]
\tikzstyle{process} = [rectangle, minimum width=2.5cm, minimum height=1cm, text centered, draw=black, fill=white]
\tikzstyle{decision} = [diamond, aspect=2, text centered, draw=black, fill=white]
\tikzstyle{data} = [trapezium, trapezium left angle=70, trapezium right angle=110, minimum width=2.5cm, minimum height=1cm, text centered, draw=black, fill=white]
\tikzstyle{database} = [cylinder, shape border rotate=90, aspect=0.25, minimum height=1cm, text centered, draw=black, fill=white]
\tikzstyle{arrow} = [thick,->,>=stealth]

\title{Augmented Reality Assistive Technologies for People with Disabilities: Enhancing Accessibility and Independence }

\author{
Riju Marwah$^{1}$\thanks{\ \ Corresponding author: \texttt{marwah.riju@gmail.com}} \quad
Jyotin Singh Thakur$^{2}$ \quad
Pranav Tanwar$^{3}$ \\
$^{1}$Guru Gobind Singh Indraprastha University, India \quad
$^{2}$Netaji Subhas University of Technology, India \\
$^{3}$Indraprastha Institute of Information Technology, India \\
}

\begin{document}
\maketitle
\begin{abstract}
Augmented Reality (AR) technologies hold immense potential for revolutionizing the way individuals with disabilities interact with the world. AR systems can provide real-time assistance and support by overlaying digital information over the physical environment based on the requirements of the use, hence addressing different types of disabilities. Through an in-depth analysis of four case studies, this paper aims to provide a comprehensive overview of the current-state-of-the-art in AR assistive technologies for individuals with disabilities, highlighting their potential to assist and transform their lives. The findings show the significance that AR has made to bridge the accessibility gap, while also discussing the challenges faced and ethical considerations associated with the implementation across the various cases. This is done through theory analysis, practical examples, and future projections that will motivate and seek to inspire further innovation in this very relevant area of exploration.
\end{abstract}

\section{Introduction}

Augmented Reality (AR) has emerged as a trans-
formative technology that overlays digital informa-
tion onto the real world to create a sort-of hybrid
environment, enhancing an individual’s perception
of reality. AR can be used in many ways such as
giving navigation guidance to a user, or generating
real-time text-to-speech for a deaf individual and
further more. Society has greatly benefited through
AR in fields like medicine, education, and assistive
technology and it further can by providing support
to the diverse needs of the population with a variety
of disabilities.

AR for the disabled is a domain of assistive tech-
nologies that can greatly increase self-sufficiency
and quality of life for many individuals. With inte-
gration of digital information into a physical envi-
ronment. AR can help with a variety of problems
faced by individuals with visual, hearing, motor
impairment and more. While promising, the AR
landscape needs further exploration and innovation,
especially in this specific domain.

This paper discusses the current state-of-the-art
in assistive AR technologies, and aims to provide a
full overview of multiple use cases of them for indi-
viduals with disabilities. Here, we take an extensive
look at how AR can can be used as assistive tech-
nology through four case studies — live captions
for deaf people, Navigation for visually impaired,
For mute people - sign language interpreter, Parkin-
son’s disease AR assistance. This paper will also
discuss the technological challenges that may be
faced and ethics of these technologies, as well as
the future endeavors which may call for innovation
and encourage research.

\section{Literature Review}

\subsection{Historic Context}
The development of assistive technologies has a long history, with early assistive devices being primarily mechanical, such as hearing aids, wheelchairs, walking sticks, etc. With the rise of digital technologies in the late 20th century, these assistive technologies expanded to text-to-speech devices, computer-based aids, and more. 

The concept of Augmented Reality (AR) emerged in the 1990s, and with the increasing availability of mobile devices such as smartphones and smartwatches, AR became widely accessible to the public. This presented great potential for exploration in assistive technologies. AR’s ability to overlay digital information onto a physical environment has opened many opportunities for developing assistive tools. This evolution from mechanical to digital assistive technology has led to AR tools that have significantly improved the quality of life for many disabled individuals.

\subsection{Current State-of-the-Art}
Current state-of-the-art AR applications for accessibility mainly focus on real-time, physically integrated data visualization with decision-making capabilities. AR provides overlays directly on the user’s surroundings, reducing cognitive load and enabling better decision-making. 

However, several challenges remain, including limited integration with decision support systems, device accessibility and costs, user experience, accuracy and reliability, and ethical concerns.

\subsection{Relevant Studies}
Several studies have demonstrated the potential of AR in assistive technologies for disabled individuals. The ARIANNA+ system, developed by Eramo et al. (2022), provides navigation guidance for the visually impaired and demonstrated the effectiveness of real-time assistance. Researchers at Rice University developed an AR application to assist patients with Parkinson’s Disease by using visual cues to overcome freezing of gait (FOG). Alam et al. (2021) created a machine-learning-based AR system to translate sign language into auditory speech, facilitating communication for deaf individuals. 

These studies highlight AR’s promise in supporting disabled individuals, although several challenges remain to be addressed.

\section{Parkinson’s Disease Assistance}

\subsection*{Introduction}
Parkinson’s Disease is a neurodengrative, progressive disorder. It is a movement disorder of the nervous system that degrades over time. As neurons weaken, get damaged or die, individuals may start noticing difficulty in movements including tremors and stiffness in legs and other parts of the body. Simple activities like walking, running and basic movements may become challenging, not all individuals with these symptoms have PD, as these appear in various other diseases as well. Traditional treatment involves medications and physical treatment, but in many cases these do not positively affect the condition of those affected. This is where AR steps in, Augmented Reality offers an approach to improve the cognitive functioning and motor control in individuals with PD.

\subsection*{Current Solutions}
Augmented Reality based innovations for Parkinson’s Disease have been created to support individuals with both motor and cognitive rehabilitation. These include smart wearables or mobile applications that will guide users through exercises and common activities. These devices and applications will provide haptic and auditory feedback. For motor rehabilitation, \ref{fig:rice-parkinson} AR can help with one of the most debilitating symptoms of Parkinson’s, the difficulty with walking, which sometimes leads to freezing of gait (FOG) which is a temporary, involuntary inability to move that can occur in individuals suffering from Parkinson’s Disease. AR system can guide individuals by projecting visual lines or steps in front of the user, helping them overcome FOG. AR can guide individuals through exercises and provide real-time feedback on their movements. This can improve muscle strength, coordination and balance, vital for the management of PD. In case of cognitive rehabilitation, memory games attention tasks can be given through AR, integrating into the real world, offering cognitive stimulation. Individuals can also be assisted with day-to-day tasks with AR by guiding them with step-by-step instructions.

\subsection*{Case Study}
An augmented reality iPhone app was designed by Rice University engineering students that would help patients with the PD symptom Freezing of Gait (FOG). For many of these patients, it was found that visual, audio or vibratory cues helped them overcome freezing. The application also allows for users to select audio cues through the mobile device’s vibration and audio capabilities. The user can select whether they want sounds, vibrations or both. It allows for customization of time between each cue. Once the cue is set and played, after every set interval a beep is heard that individuals can follow with their steps that will prevent them from freezing. This app is a great way for patients to overcome many symptoms of Parkinson’s Disease. The application utilizes augmented reality where the user can point the mobile device towards the ground and trigger it to place a box or circle where the foot of the patient should land. Once placed, the app places another where the foot should land next. The visual cue often allows users to overcome FOG.

\begin{figure}[t]
    \centering
    \includegraphics[width=\linewidth]{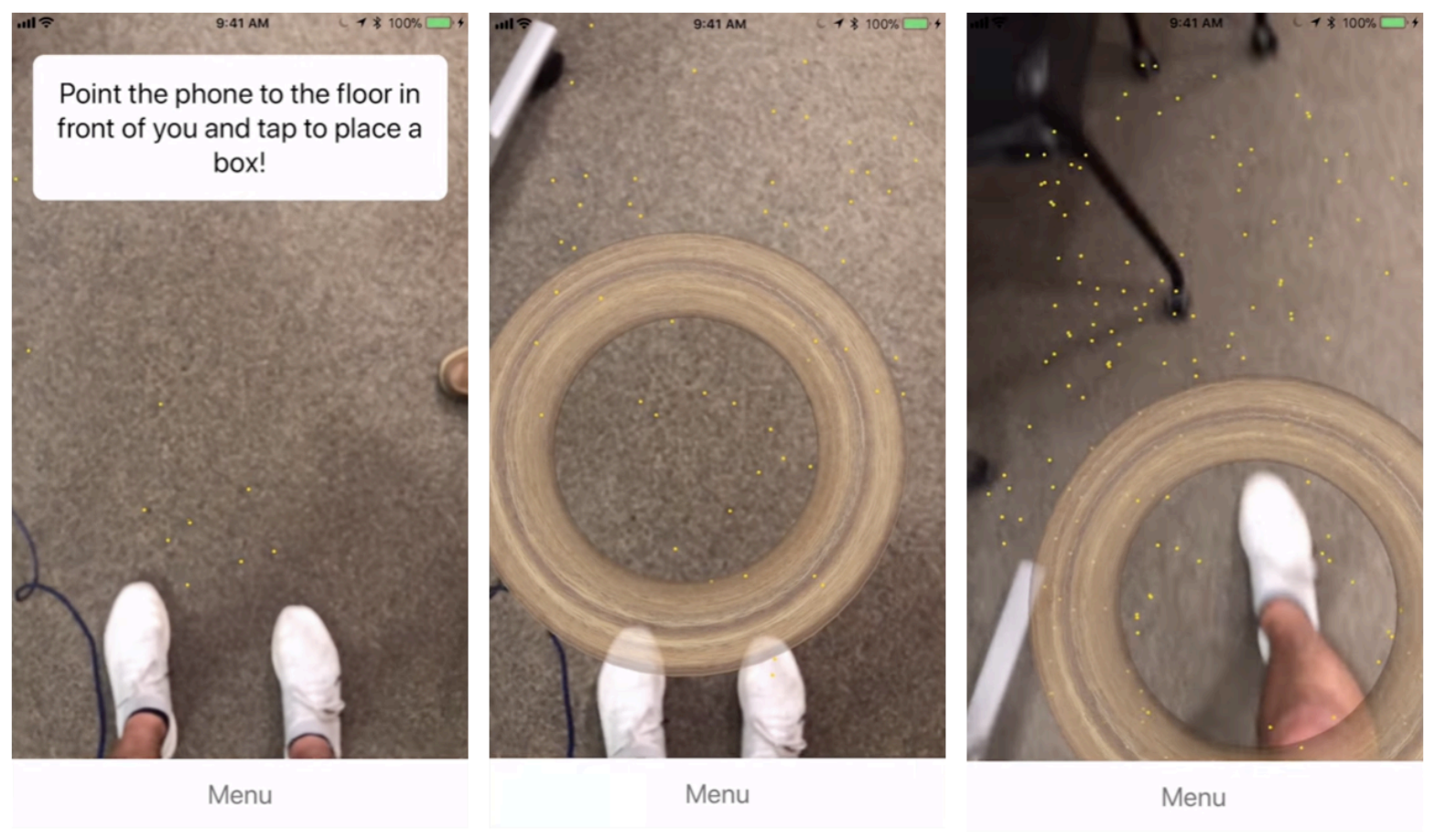}
    \caption{Rice University — \emph{Augmented reality app may aid patients with Parkinson's}. 
    App allows the user to choose between different shapes and point towards the floor, then tap to place the shape; The shape (ring) is placed where the user should step, and continues to be placed ahead of the user to guide movement.}
    \label{fig:rice-parkinson}
\end{figure}

The object placement in the application works by \ref{fig:rice-parkinson} using AR to place virtual objects, such as boxes, rings or pathways, within the user’s real surroundings. The objects act as targets where the user is supposed to step next to initiate frozen movement. Real-time mapping is done by utilizing the smartphone’s camera system, allowing the application to dynamically place these targets wherever needed based on the location and movements. The app allows for customization of objects placed on the user’s display to allow for more visual stimulation. (Rice University, 2019)

\subsection*{Conclusion}
Studies have shown that AR-based gait training can lead to major improvements in affected individual’s movements including stability, confidence, walking speed, etc. The real-time feedback of the app single-handedly shows more improvements than classic, traditional physical therapy.

\section{Navigation for the Visually Impaired}

\subsection*{Introduction}
Out of the variety of problems faced by visually impaired individuals, navigating through physical spaces autonomously is the most important. Mechanical aids such as walking sticks and guide dogs have limited usability, particularly in environments of complex nature. Augmented reality bridges this gap by giving individuals wear real-time wearable navigation aids that guide them to move around physical spaces in a safe manner.

\subsection*{Current Solutions}
There have been several developments of AR-bases systems that have been created to aid visually impaired individuals, these usually involve a form such as wearables like smart glasses or mobile devices which run AR applications that return audio or haptic feedback. Some popular examples are object and obstacle recognition within travel path of the user of the device, which it then relates to the user through audio or haptic feedback. Another scenario is a navigation system to direct the user precisely through audio feedback.

\begin{figure}[t]
    \centering
    \includegraphics[width=\linewidth]{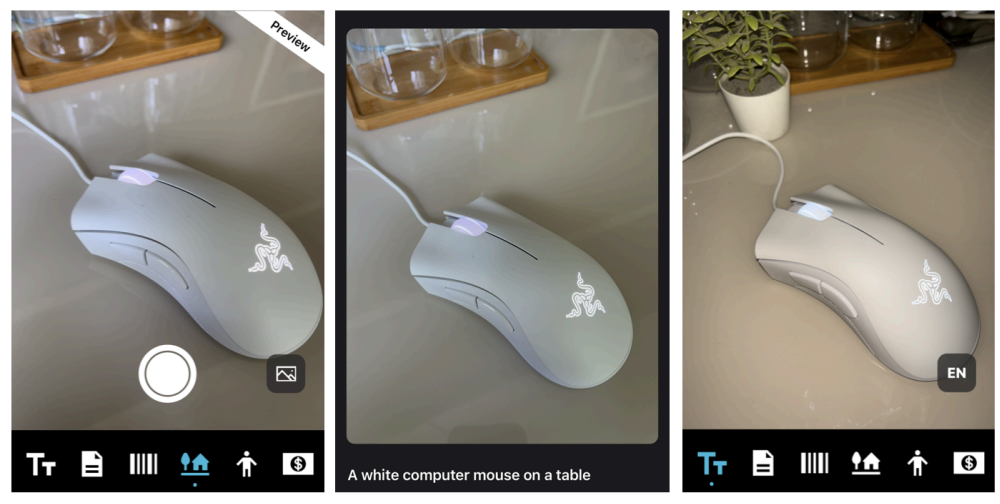}
    \caption{Preview of Microsoft’s \emph{Seeing AI} application. 
    Upon first opening, the application presents the camera interface; Capturing an image of the environment returns both a visual and audio description of the scene; In low-light conditions, the smartphone’s flash activates to assist object recognition.}
    \label{fig:seeing-ai}
\end{figure}

Microsoft developed an application “Seeing AI”, which uses AR features to detect objects within the frame of an image taken through the app and describes the scene through audio. This application is easily accessible and available on iOS and Android devices. First opening the “Seeing AI” application, a camera output is presented with several options at the bottom of the frame. These options present various methods in which the application can detect text, environments, people, documents, and more. Using the environment option as a test case, a picture is taken and the scene is correctly described in both visual and audio responses — “A white computer mouse on a table”. It was tested that the application is highly accurate in describing scenes and usually handles most details without any issues.

% Figure placeholder for Seeing AI Object Recognition Workflow (Figure 7)

\subsection{Case Study: Seeing AI and ARIANNA+}

The core of Seeing AI’s ability to recognize and identify objects is the AI and ML models it uses. Here is an overview of how this process takes place:

\begin{itemize}
    \item \textbf{Image Acquisition:} The process begins when the application captures an image; this visual data is used as the input for the AI and ML models.
    \item \textbf{Image Preprocessing:} Before the image is used by the models, it is first preprocessed to optimize it for further analysis, including cropping, color correction, and normalizing pixel values. Computer vision techniques such as edge detection may be used to recognize contours of objects, aiding identification.
    \item \textbf{Convolutional Neural Networks (CNNs):} Seeing AI likely uses CNNs—deep learning models specifically designed for image recognition.  
    Layers of CNNs include:
    \begin{itemize}
        \item \emph{Convolutional Layers:} Apply filters (kernels) scanning for low-level features such as edges, corners, and textures.
        \item \emph{Pooling Layers:} Reduce spatial dimensions of feature maps, lowering computational load and improving efficiency.
        \item \emph{Activation Functions:} Non-linear activations such as ReLU enable learning of complex patterns.
        \item \emph{Fully Connected Layers:} Combine extracted features for final classification decisions.
    \end{itemize}
    \item \textbf{Object Classification:} The final CNN layer is often a softmax layer, outputting a probability distribution over object classes and returning the top-$k$ predictions.
    \item \textbf{Transfer Learning:} Pre-trained CNNs (e.g., Inception, MobileNet, ResNet) are fine-tuned on domain-specific datasets relevant to daily environments.
    \item \textbf{Object Detection \& Localization:} Region-based CNNs (R-CNNs) or YOLO are used to detect and localize objects, drawing bounding boxes; Mask R-CNN may be used for instance segmentation.
    \item \textbf{Contextual Understanding:} Models analyze relationships between objects and may integrate GPS, gyroscope, and other sensor data for enhanced scene understanding.
    \item \textbf{Continuous Learning:} Models are updated with new training data to improve recognition capabilities over time.
    \item \textbf{Natural Language Processing (NLP):} Classification results are converted to natural language descriptions, which are then spoken aloud via text-to-speech (TTS).
\end{itemize}

The \textbf{ARIANNA+ System} presents an AR-based solution to guide visually impaired individuals in both indoor and outdoor environments, creating visual paths using AR and CNNs that can be followed on a smartphone. Building upon the original ARIANNA system’s tactile paths, ARIANNA+ instead projects virtual navigation paths and uses CNNs for obstacle detection. Haptic and auditory feedback guide users in real-time.

The ARIANNA+ system integrates multiple technologies as follows:
\begin{itemize}
    \item \textbf{Augmented Reality (AR):} Creates virtual paths overlaid on the live camera view.
    \item \textbf{Convolutional Neural Networks (CNNs):} Detect obstacles and relevant environmental features.
    \item \textbf{Smartphone Integration:} Utilizes onboard cameras and sensors to capture surroundings and track movement for real-time AR rendering.
    \item \textbf{Feedback Mechanisms:} Provides auditory and haptic feedback for navigation cues.
\end{itemize}

Further studies examine urban navigation systems for visually impaired users, including smartphones, wearables, and sensor-based solutions for object detection, environmental analysis, and route planning. Others review the shift from traditional aids to electronic travel aids (ETAs), noting that despite technological advances, challenges remain in usability, multimodal feedback integration, and interface design.

\begin{table}[h!]
\centering
\resizebox{0.5\textwidth}{!}{ % 85% of text width
\begin{tabular}{l c c c}
\toprule
\textbf{Application} & \textbf{Object Recognition Accuracy} & \textbf{Scene Description} \\
\midrule
Seeing AI  & 95\% & 90\% \\
ARIANNA+  & 93\% & N/A \\
Other ETAs & Varies (85–95\%) & N/A \\
\bottomrule
\end{tabular}
}
\caption{Accuracy and Performance Metrics}
\label{tab:accuracy-performance}
\end{table}

\begin{table*}[h!]
\centering
\resizebox{0.85\textwidth}{!}{ % 95% of full width
\begin{tabular}{l c c c c c}
\toprule
\textbf{Feature} & \textbf{Seeing AI} & \textbf{ARIANNA+} & \textbf{Other ETAs} \\
\midrule
Object Detection       & Yes (CNNs, R-CNNs, YOLO) & Yes (CNNs)       & Yes (various methods) \\
Augmented Reality      & No                       & Yes              & Limited              \\
Real-time Feedback     & Yes (Audio via TTS)       & Yes (Audio, Haptic) & Yes (varied)        \\
GPS Integration        & Possibly                 & Yes              & Yes                  \\
Environmental Analysis & Yes                      & Yes              & Yes                  \\
Usability              & High                     & High             & Varies                \\
\bottomrule
\end{tabular}
}
\caption{Comparative Analysis of Assistive Applications}
\label{tab:assistive-apps}
\end{table*}

\begin{figure}[h!]
    \centering
    \begin{tikzpicture}
        \begin{axis}[
            width=0.85\linewidth,
            height=0.6\linewidth,
            xlabel={Processing Speed (ms)},
            ylabel={Accuracy (\%)},
            ymin=78, ymax=98,
            xmin=0, xmax=420,
            grid=major,
            legend style={at={(0.5,-0.15)},anchor=north,legend columns=1},
            legend cell align=left,
            tick label style={/pgf/number format/fixed},
            ylabel near ticks,
            xlabel near ticks
        ]
        
        % CNN-based Models
        \addplot[
            color=blue,
            mark=*,
            thick
        ] coordinates {
            (50,95) (100,97) (150,97) (200,97)
        };
        \addlegendentry{Technology A (CNN-based Models)}
        
        % YOLO Models
        \addplot[
            color=orange,
            mark=*,
            thick
        ] coordinates {
            (10,80) (25,85) (50,90)
        };
        \addlegendentry{Technology B (YOLO Models)}
        
        % R-CNN Models
        \addplot[
            color=green!70!black,
            mark=*,
            thick
        ] coordinates {
            (200,90) (300,93) (350,93) (400,93)
        };
        \addlegendentry{Technology C (R-CNN Models)}

        \end{axis}
    \end{tikzpicture}
    \caption{Processing Speed vs. Accuracy Trade-off}
    \label{fig:processing-speed-accuracy}
\end{figure}

\subsubsection{Challenges}

\begin{itemize}
    \item \textbf{Customization \& Adaptability:} Current AR systems lack flexibility to adapt to individual needs.
    \item \textbf{Latency Issues:} High latency in image processing can hinder real-time obstacle detection.
    \item \textbf{Multimodal Feedback:} Integrating haptic, auditory, and other sensory cues effectively remains challenging.
    \item \textbf{Usability \& Accessibility:} Interfaces must accommodate a variety of sensory input modes.
    \item \textbf{Accuracy \& Reliability:} Misidentifications can lead to unsafe situations.
    \item \textbf{Environmental Variability:} Poor lighting or complex surroundings can reduce performance.
    \item \textbf{Contextual Awareness:} Systems may fail to distinguish relevant from irrelevant contextual cues.
    \item \textbf{Data Privacy \& Security:} Large training datasets raise sensitive data concerns.
\end{itemize}

\subsubsection{Proposed Solutions}

\begin{itemize}
    \item AI-driven interfaces that adapt layouts and feedback types in real-time.
    \item Use of 5G and edge computing to reduce latency.
    \item Development of universal frameworks with multi-platform support.
    \item Integration of additional sensors (LiDAR, infrared) for enhanced accuracy.
    \item Context-aware engines combining IoT sensor and GPS data.
    \item Augmenting physical mobility aids with AR modules for seamless integration.
    \item Environment-aware feedback adjustment (e.g., enabling flash in low light).
\end{itemize}

\subsubsection{Conclusion and Future Directions}

This study explored how AR can enable autonomous navigation for visually impaired individuals. Integrating AR with AI and ML can bridge accessibility gaps, providing real-time guidance without physical aids. Future work should focus on improved customization, more precise object recognition, and affordable solutions for wider adoption.

\subsection{Live Captions for the Deaf}
\textbf{Introduction:} Deaf individuals face a fundamental communication challenge, as they primarily rely on hand gestures and body language rather than speech. Most hearing individuals do not understand sign language, creating a need for a translator capable of interacting with both deaf and non-deaf individuals. This translator must be proficient in the specific sign language system (e.g., American Sign Language, Indian Sign Language) used by the deaf person and in the spoken language of the hearing person. The existence of various sign language systems also necessitates translation between two deaf individuals using different systems. Without such knowledge or a translator, efficient communication becomes difficult.

\begin{figure}[t]
    \centering
    \includegraphics[width=\linewidth]{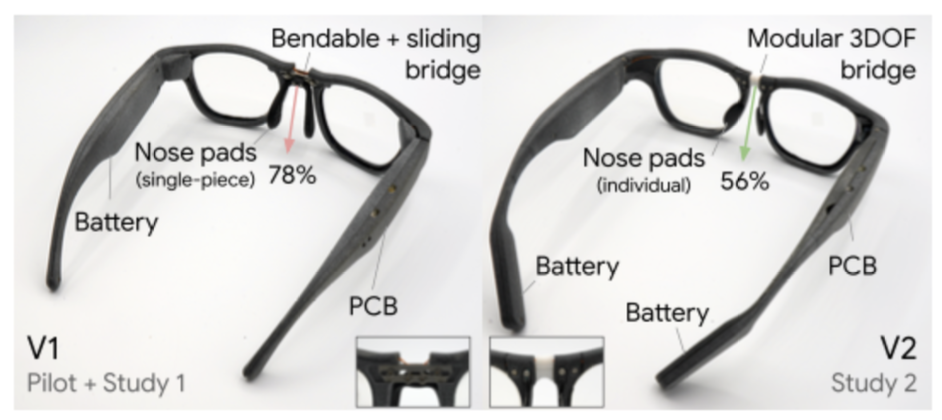}
    \caption{HWD Sunglasses.}
    \label{fig:hwd-sunglasses}
\end{figure}

\begin{figure}[t]
    \centering
    \includegraphics[width=\linewidth]{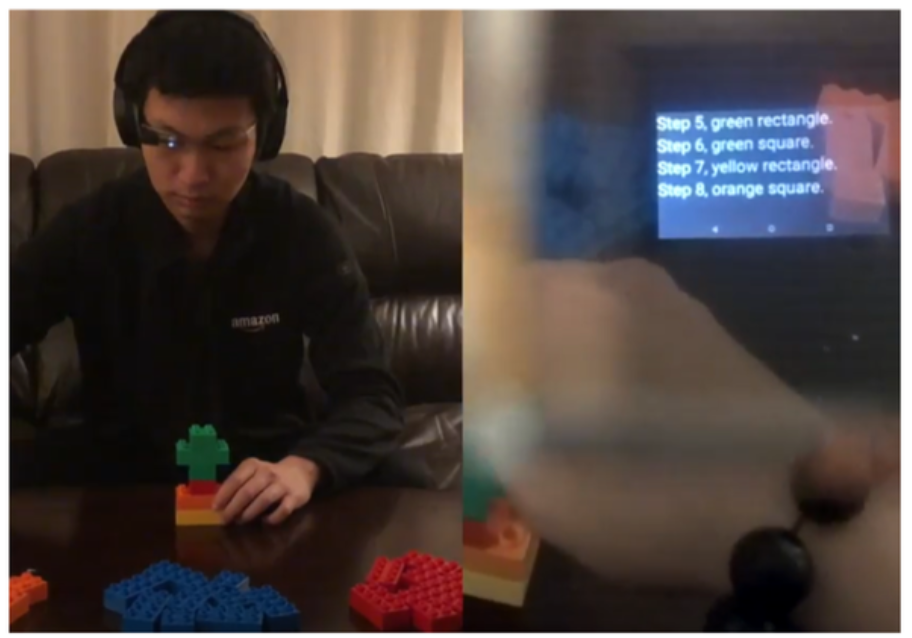}
    \caption{Real-time Captioning in Google Glasses.}
    \label{fig:google-sunglasses}
\end{figure}

\textbf{Current Solutions and Technology:} Current solutions involve Augmented Reality glasses that display live subtitles for spoken text~\cite{ref1}. However, challenges include discomfort due to heavy eyewear, faded or illegible captions, and increased eye strain~\cite{ref2}. Some studies found that phone-based translator applications were more comfortable, though they introduced cognitive load from holding the phone, reading captions, and replying. While Head-Mounted Displays (HMDs) can address captioning issues, their bulkiness causes discomfort, limiting user preference. Advancements such as visualizing sound intensity have been explored~\cite{ref6}, but two-way translation between speech and sign language remains rare. 

Some systems translate sign language to text with higher accuracy than HWDs~\cite{ref3}, while others tackle multiple sign language translation~\cite{ref4}, showing advantages over separate bilingual systems. Behavioral analysis using gesture detection frameworks (ABA-GDF)~\cite{ref5} also demonstrated higher accuracy than conventional methods.

\begin{table*}[h!]
\centering
\resizebox{0.7\textwidth}{!}{ % 90% of full width
\begin{tabular}{c c c c c c}
\toprule
\textbf{Language Pairs} & \multicolumn{2}{c}{\textbf{Single MLSLT}} & \multicolumn{2}{c}{\textbf{BSLT}} & \textbf{Param ratio} \\
\cmidrule(lr){2-3} \cmidrule(lr){4-5}
 & BLEU4 & ROUGE & BLEU4 & ROUGE &  \\
\midrule
2 × 2   & 2.43 & 30.05 & 3.62 & 33.76 & 4:1   \\
3 × 3   & 2.53 & 30.54 & 3.34 & 32.72 & 9:1   \\
4 × 4   & 2.24 & 29.55 & 2.72 & 31.69 & 16:1  \\
5 × 5   & 2.37 & 29.00 & 2.81 & 31.25 & 25:1  \\
6 × 6   & 2.26 & 30.01 & 3.05 & 31.58 & 36:1  \\
7 × 7   & 2.02 & 29.17 & 2.49 & 29.96 & 49:1  \\
8 × 8   & 1.78 & 28.87 & 2.04 & 30.00 & 64:1  \\
9 × 9   & 1.71 & 28.92 & 2.09 & 30.01 & 81:1  \\
10 × 10 & 1.66 & 29.27 & 1.88 & 30.26 & 100:1 \\
\bottomrule
\end{tabular}
}
\caption{Comparison of MLSLT and BSLT on various language pairs.}
\label{tab:mlslt-bslt}
\end{table*}

\begin{figure}[h!]
\centering
\resizebox{0.45\textwidth}{!}{ % 80% of text width
\begin{tikzpicture}
\begin{axis}[
    ybar,
    bar width=7pt,
    enlarge x limits=0.15,
    ylabel={Performance Analysis Ratio (\%)},
    xlabel={Number of Samples},
    ymin=0, ymax=100,
    xtick=data,
    xticklabels={10,20,30,40,50,60,70,80,90,100},
    legend style={at={(0.5,-0.15)}, anchor=north, legend columns=-1},
    ymajorgrids=true,
    grid style=dashed,
]
\addplot[fill=yellow!50!brown] coordinates {(1,64) (2,66) (3,68) (4,69) (5,70) (6,71) (7,71) (8,72) (9,73) (10,74)};
\addplot[fill=purple!70!black] coordinates {(1,38) (2,15) (3,43) (4,34) (5,58) (6,59) (7,42) (8,70) (9,73) (10,85)};
\addplot[fill=green!70!black] coordinates {(1,53) (2,54) (3,55) (4,57) (5,58) (6,59) (7,61) (8,62) (9,64) (10,90)};
\addplot[fill=red!70!black] coordinates {(1,58) (2,43) (3,68) (4,58) (5,78) (6,60) (7,77) (8,86) (9,89) (10,92)};
\legend{RCNN, HGR-CV, FDOSM, ABA-GDF}
\end{axis}
\end{tikzpicture}
} % end resizebox
\caption{Graph comparing traditional translation methods and ABA-GDF.}
\label{fig:aba-gdf}
\end{figure}

\textbf{Conclusion:} Most current solutions address isolated aspects such as translation accuracy, ease of use, or multi-language support, but few integrate them into a unified AR-based system. Translation between different sign language systems also remains underexplored.

\textbf{Future Directions:} A comprehensive solution should combine multiple-language translation, sign-to-sign translation, sound intensity visualization, ABA-GDF, and HoloSound~\cite{ref6} into a single device. Improving HWD comfort, affordability, and quality is critical. Involving Deaf and Hard of Hearing (DHH) individuals in development could ensure better contextual awareness. Cost reduction is also vital to make these solutions accessible, particularly in low-income regions.

\subsection{Assistance for Mute People}
\textbf{Introduction:} Mute individuals often face difficulties communicating since voice is a primary means of interaction. Most hearing individuals do not understand sign language, necessitating pen-and-paper methods or a translator. Augmented reality, machine learning-based sign recognition, and integration into everyday devices can help bridge this gap.

\textbf{Challenges:} Communication barriers occur when hearing individuals do not know sign language and mute individuals do not understand the spoken language. This mismatch necessitates translation tools that work bidirectionally. Existing solutions are often cumbersome, insecure, or slow.

\textbf{Current Solutions and Technology:} AR glasses under development by Google and Meta can convert sign language into text using ML algorithms, displaying captions directly to the hearing individual~\cite{ref11}. However, they may be error-prone in low-light environments. Another approach uses gloves embedded with flex sensors to translate signs into audio~\cite{ref12,ref13,ref14}. These gloves may also include displays, but high sensor costs and sign language’s reliance on facial expressions make them incomplete solutions.

\begin{figure}[h!]
\centering
\scalebox{0.3}{ % Adjust size
\begin{tikzpicture}[node distance=2cm]

% Styles
\tikzstyle{startstop} = [ellipse, minimum width=2cm, minimum height=1cm,text centered, draw=black, fill=white]
\tikzstyle{process} = [rectangle, minimum width=2.5cm, minimum height=1cm, text centered, draw=black, fill=white]
\tikzstyle{decision} = [diamond, aspect=2, text centered, draw=black, fill=white]
\tikzstyle{data} = [trapezium, trapezium left angle=70, trapezium right angle=110, minimum width=2.5cm, minimum height=1cm, text centered, draw=black, fill=white]
\tikzstyle{database} = [cylinder, shape border rotate=90, aspect=0.25, minimum height=1cm, text centered, draw=black, fill=white]
\tikzstyle{arrow} = [thick,->,>=stealth]

% Nodes
\node (start) [startstop] {Start};
\node (handgesture) [process, right of=start, xshift=3.5cm] {Hand Gesture};
\node (sensor) [data, right of=handgesture, xshift=4cm] {Sensor};
\node (sign) [decision, right of=sensor, xshift=4cm] {Sign Recognized};
\node (dict) [database, above of=sign, yshift=2.5cm] {Dictionary};

% New vertical flow after decision node
\node (speech) [data, below of=sign, yshift=-2.5cm] {Speech Synthesis};
\node (output) [process, below of=speech, yshift=-2.5cm] {Audio+Video Output};
\node (end) [startstop, below of=output, yshift=-2.5cm] {End};

% Arrows
\draw [arrow] (start) -- (handgesture);
\draw [arrow] (handgesture) -- (sensor);
\draw [arrow] (sensor) -- (sign);
\draw [arrow] (sign) -- (speech);
\draw [arrow] (speech) -- (output);
\draw [arrow] (output) -- (end);
\draw [arrow] (sign) -- (dict);
\draw [arrow] (dict) -- (handgesture);

\end{tikzpicture}
} % End of scale
\caption{Working sequence of glove-based solution.}
\label{fig:glove-sequence}
\end{figure}

\begin{table}[h!]
\centering
\begin{tabular}{lcc}
\hline
\textbf{Logics} & \textbf{Response Time (s)} & \textbf{Accuracy (\%)} \\
\hline
1 & 2 & 98 \\
2 & 1.3 & 98 \\
3 & 1 & 98 \\
4 & 2 & 98 \\
5 & 1 & 94 \\
6 & 1.5 & 94 \\
7 & 2.3 & 95 \\
8 & 2 & 93 \\
\textbf{Average} & \textbf{1.63} & \textbf{96} \\
\hline
\end{tabular}
\caption{Accuracy of the device for the logics [Speaking System for Deaf and Mute People with Flex Sensors]}
\label{tab:accuracy-logics}
\end{table}

\textbf{Conclusion:} While existing solutions enable communication for mute individuals, they are still in early stages and often hardware-dependent.

\textbf{Future Directions:} A promising approach is a mobile application that translates spoken language into both text and audio understood by the mute person, and vice versa. Integration with AR glasses (e.g., Meta Ray-Ban) could allow real-time captions and auditory output. Such a system would ease communication between all combinations of mute, deaf, and hearing individuals, provided that cost and usability challenges are addressed.

\begin{table}[h!]
\centering
\scriptsize
\begin{tabular}{c c c c c l}
\toprule
\textbf{Logic} & \multicolumn{4}{c}{\textbf{Four Finger Logic}} & \textbf{Actions} \\
\midrule
Logic 1 & 1 & 0 & 0 & 0 & I feel danger. I need your help. \\
Logic 2 & 0 & 1 & 0 & 0 & I want to drink. \\
Logic 3 & 0 & 0 & 1 & 0 & Thank you. \\
Logic 4 & 0 & 0 & 0 & 1 & Sending SMS. \\
Logic 5 & 1 & 1 & 0 & 0 & Speak Up. \\
Logic 6 & 1 & 1 & 1 & 0 & Where is my medicine? \\
Logic 7 & 0 & 1 & 1 & 0 & Can you hear me? If yes, come to me. \\
Logic 8 & 0 & 1 & 1 & 1 & Please turn OFF the lights. \\
\bottomrule
\end{tabular}
\caption{Logic and implemented actions of the device [Speaking System for Deaf and Mute People with Flex Sensors]}
\label{tab:logic-actions}
\end{table}

\section{Theoretical Analysis}
Since AR adds digital content to the environment, individuals are able to gain a greater sense of their surroundings. This integration enables timely decision-making, minimizing cognitive overload on disabled individuals through quick feedback. In the context of navigation for the visually impaired, AR can provide auditory and haptic feedback to assist with cognitive load reduction and enhance spatial awareness. 

Applications also extend to Parkinson’s Disease, where projection of visual cues—either from wearable devices or mobile applications—helps restore motor control by alleviating freezing of gait (FOG). The development of such systems draws on human-computer interaction principles, aiming to reduce cognitive load and improve quality of life. Despite current challenges, the potential for AR in assistive technology provides numerous opportunities for advancing theoretical models for accessibility-focused solutions.

\section{Future Projections}
The future of AR assistive technologies for individuals with disabilities is highly promising. In the coming years, we anticipate a shift toward more individualized and adaptable AR systems that cater to user-specific needs. The integration of artificial neural networks (ANNs) into AR will allow users to fine-tune preferences, enabling personalized and context-aware experiences.

Significant advancements are expected in multimodal feedback, combining visual, haptic, and auditory cues to enhance interaction. AR systems will also evolve toward greater environmental awareness and object recognition, further reducing cognitive load. Additionally, decreasing device costs will expand accessibility, making these tools available to a wider population. As adoption grows, so too will the role of AR in improving independence and accessibility for individuals with disabilities.

\section{Conclusion}
The use of AR in assistive technologies has demonstrated substantial benefits in improving accessibility and independence for people with disabilities. By merging real-world environments with digital overlays, AR systems offer real-time feedback and context awareness, enriching user experiences. However, challenges remain in achieving full customization, flexibility, and low-latency operation.

Ongoing research and innovation will be key to addressing these limitations. In the coming years, with advances in affordability, personalized interfaces, and integrated multimodal feedback, individuals with disabilities will be able to fully experience the potential of AR. This paper emphasizes the need for continued exploration of innovative, inclusive solutions that enhance quality of life for all.


\begin{thebibliography}{99}
\bibitem{ref1} Olwal, A., Balke, K., Votintcev, D., Starner, T., Conn, P., Chinh, B., \& Corda, B. (2020). Wearable subtitles: Augmenting spoken communication with lightweight eyewear for all-day captioning. \textit{Proceedings of the 33rd Annual ACM Symposium on User Interface Software and Technology (UIST '20)}, 20–23. ACM. \url{https://doi.org/10.1145/3429301.3430358}

\bibitem{ref2} Tu, J., Lin, G., \& Starner, T. E. (2020). Towards an understanding of real-time captioning on head-worn displays. \textit{Proceedings of the 22nd International Conference on Human-Computer Interaction with Mobile Devices and Services (MobileHCI '20 Extended Abstracts)}. ACM. \url{https://doi.org/10.1145/3406324.3410725}

\bibitem{ref3} Alam, F. R., Munir, M. B., Ishrak, S., Hussain, S., Shalahuddin, M., \& Islam, M. N. (2021). A machine learning-based sign language interpretation system for communication with deaf-mute people. \textit{Proceedings of the 21st International Conference on Human Computer Interaction (Interacción 2021)}. ACM. \url{https://doi.org/10.1145/3476081.3476191}

\bibitem{ref4} Yin, A., Zhao, Z., Jin, W., Zhang, M., Zeng, X., \& He, X. (2022). MLSLT: Towards multilingual sign language translation. \textit{Proceedings of the IEEE/CVF Conference on Computer Vision and Pattern Recognition (CVPR)}. IEEE. \url{https://doi.org/10.1109/CVPR52688.2022.00034}

\bibitem{ref5} Nirmala, M. S. (2023). Behavioral analysis of deaf and mute people using gesture detection. \textit{Proceedings of the 23rd International Conference on Human-Computer Interaction (HCI 2023)}. ACM. \url{https://doi.org/10.1145/3509899.3510000}

\bibitem{ref6} Jain, D. (2022). Sound sensing and feedback techniques for deaf and hard of hearing people. \textit{International Journal of Human-Computer Interaction}, 38(12), 1107–1118. \url{https://doi.org/10.1080/10447318.2021.1996112}

\bibitem{ref7} Eramo, V., Chiaravalloti, F., Giampetruzzi, E., Baldoni, R., \& Pannone, D. (2022). ARIANNA+: A new augmented reality application for navigation of people with visual impairments. \textit{IEEE Transactions on Human-Machine Systems}, 52(3), 291–304. \url{https://doi.org/10.1109/THMS.2022.3159374}

\bibitem{ref8} Messaoudi, M. D., Menelas, B. A. J., \& Mcheick, H. (2022). Review of navigation assistive tools and technologies for the visually impaired. \textit{Sensors}, 22(20), 7888. \url{https://doi.org/10.3390/s22207888}

\bibitem{ref9} Nair, A. B., Shinde, A. R., \& Patil, S. A. (2023). Urban navigation systems for visually impaired people: A systematic review. \textit{Journal of Assistive Technologies}, 17(1), 45–58. \url{https://doi.org/10.1108/JAT-11-2022-0045}

\bibitem{ref10} Rice University. (2019, May 1). Augmented reality app may aid patients with Parkinson's [Video]. YouTube. \url{https://www.youtube.com/watch?v=me6mnnn89h4}

\bibitem{ref11} Mehra, V., Pandey, D., Rastogi, A., Singh, A., \& Singh, H. P. (2021). Technological Aids for Deaf and Mute in the Modern World. \textit{Recent Patents on Engineering}, 15(6), 43–52. \url{https://doi.org/10.2174/1872212114999201116214802}

\bibitem{ref12} Istiqomah, K. N., Alditama, R., Salsabila, S., \& ‘Azzam, A. (2023). Development of communication tools for deaf and mute people using design thinking method. \textit{AIP Conference Proceedings}, 2828, 060001. \url{https://doi.org/10.1063/5.0164169}

\bibitem{ref13} Babour, A., Bitar, H., Alzamzami, O., Alahmadi, D., Barsheed, A., Alghamdi, A., \& Almshjary, H. (2023). Intelligent gloves: An IT intervention for deaf-mute people. \textit{Journal of Intelligent Systems}. \url{https://doi.org/10.1515/jisys-2022-0076}

\bibitem{ref14} Patwary, A. S., Zaohar, Z., Sornaly, A. A., \& Khan, R. (2022). Speaking System for Deaf and Mute People with Flex Sensors. \textit{2022 6th International Conference on Trends in Electronics and Informatics (ICOEI)}, 168–173. IEEE. \url{https://doi.org/10.1109/ICOEI53556.2022.9777226}
\end{thebibliography}
\end{document}